\documentclass[reqno,11pt]{amsart}
\usepackage{amssymb,amsmath}
\usepackage{color}
\usepackage{accents}
\usepackage{texdraw}
\usepackage{eucal}
\usepackage{epic,epsfig}
\usepackage{graphicx}
\usepackage[all]{xy}

\numberwithin{equation}{section}
\DeclareMathAccent{\wtilde}{\mathord}{largesymbols}{"65}
\DeclareMathAccent{\what}{\mathord}{largesymbols}{"62}

\def\m@th{\mathsurround=0pt}
\mathchardef\bracell="0365
\def\upbrall{$\m@th\bracell$}
\def\undertilde#1{\mathop{\vtop{\ialign{##\crcr
    $\hfil\displaystyle{#1}\hfil$\crcr
     \noalign
     {\kern1.5pt\nointerlineskip}
     \upbrall\crcr\noalign{\kern1pt
   }}}}\limits}

\newcommand{\wb}[1]{\overline{#1}}

\newcommand{\wh}{\widehat}
\newcommand{\wt}{\widetilde}

\newcommand{\Ld}{\boldsymbol{\Lambda}}
\newcommand{\tLd}{\,^{t\!}\boldsymbol{\Lambda}}

\newcommand{\bun}{\boldsymbol{1}}
\newcommand{\bOm}{\boldsymbol{\Omega}}
\newcommand{\buk}{\boldsymbol{u}_k}

\newcommand{\tbc}{\,^{t\!}\boldsymbol{c}}

\newcommand{\tbs}{\,^{t\!}{\boldsymbol{s}}}
\newcommand{\tbme}{\,^{t\!}{\boldsymbol{e}}}

\newcommand{\bblu}{\begin{color}{blue}}
\newcommand{\bred}{\begin{color}{red}}
\newcommand{\ecl}{\end{color}}

\newcommand{\bC}{\boldsymbol{C}}

\newcommand{\bK}{\boldsymbol{K}}
\newcommand{\bL}{\boldsymbol{L}}
\newcommand{\bM}{\boldsymbol{M}}

\newcommand{\bO}{\boldsymbol{O}}

\newcommand{\bU}{\boldsymbol{U}}

\newcommand{\bb}{\beta}

\newcommand{\sg}{\sigma}

\newcommand{\ld}{\lambda}

\newcommand{\oa}{\omega}
\newcommand{\be}{\begin{equation}}
\newcommand{\ee}{\end{equation}}
\newcommand{\bea}{\begin{eqnarray}}
\newcommand{\eea}{\end{eqnarray}}
\newcommand{\bse}{\begin{subequations}}
\newcommand{\ese}{\end{subequations}}
\newcommand{\nn}{\nonumber}

\newcommand{\ol}{\overline}

\newcommand{\bu}{\boldsymbol{u}}

\newcommand{\bc}{\boldsymbol{c}}

\newcommand{\bme}{\boldsymbol{e}}

\newcommand{\brr}{\boldsymbol{r}}

\newcommand{\bphi}{{\boldsymbol \phi}}
\newcommand{\bpsi}{{\boldsymbol \psi}}

\begin{document}
\title[ Direct Linearization of extended lattice BSQ systems ]
{ Direct Linearization of extended lattice BSQ systems}
\email{djzhang@staff.shu.edu.cn}

\author{Da-jun Zhang \& Song-lin Zhao}
\address{Department of Mathematics, University of Shanghai \\ Shanghai 200444\\ China}

\author{Frank W Nijhoff}
\address{
School of Mathematics\\
University of Leeds\\
Leeds LS2 9JT\\
United Kingdom} \email{Corresponding author.
nijhoff@maths.leeds.ac.uk}

\begin{abstract}
The direct linearization structure is presented of a ``mild'' but significant
generalization of the lattice BSQ system. Some of the equations in
this system were recently discovered in [J. Hietarinta, J. Phys {\bf
A}: Math. Theor. {\bf 44} (2011) 165204] through a search of a class
of three-component systems obeying the property of multidimensional
consistency. We show that all the novel equations arising in this
class follow from one and the same underlying structure. Lax pairs
for these systems are derived and explicit expressions for the
$N$-soliton solutions are obtained from the given structure.
\end{abstract}

\maketitle

\section{Introduction}
\setcounter{equation}{0}

The lattice Boussinesq (BSQ) equation
\begin{eqnarray}
& &\frac{p^3-q^3}{p-q+u_{n+1,m+1}-u_{n+2,m}}\,-\,
\frac{p^3-q^3}{p-q+u_{n,m+2}-u_{n+1,m+1}}   \nn \\
&&\qquad = (p-q+u_{n+1,m+2}-u_{n+2,m+1})(2p+q+u_{n,m+1}-u_{n+2,m+2})  \nn \\
&&\qquad\quad- (p-q+u_{n,m+1}-u_{n+1,m})(2p+q+u_{n,m}-u_{n+2,m+1}) \
, \label{eq:dBSQ}
\end{eqnarray}
appeared in \cite{GD} as a first higher-rank case in what was coined
the ``lattice Gel'fand-Dikii (GD)'' hierarchy of equations. These
are equations on the space-time lattice, labelled by an integer $N$
associated with the order (or equivalently the number of effective
components) of the system. Whilst for $N=2$ the equation in this
class is the lattice potential Korteweg-de Vries (KdV) equation,
which is an equation associated with a quadrilateral stencil of
vertices, the lattice BSQ appears in this class for $N=3$ and is
defined on the following 9-point stencil: \vspace{.9cm}

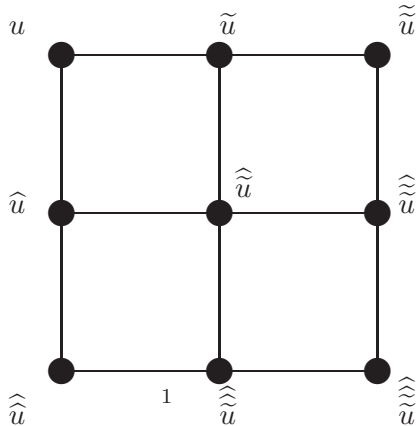
\begin{center}

\setlength{\unitlength}{.7mm}
\begin{picture}(160,60)(-60,0)
\put(0,0){\circle*{5}} \put(-10,-10){${\large{\wh{\wh{u}}}}$}
\put(0,30){\circle*{5}} \put(-10,30){${\large{\wh{u}}}$}
\put(0,60){\circle*{5}} \put(-10,64){${\large{u}}$}
\put(30,0){\circle*{5}} \put(30,-10){${\large{\wh{\wh{\wt{u}}}}}$}
\put(30,30){\circle*{5}} \put(33,33){${\large{\wh{\wt{u}}}}$}
\put(30,60){\circle*{5}} \put(30,64){${\large{\wt{u}}}$}
\put(60,0){\circle*{5}}
\put(64,-10){${\large{\wh{\wh{\wt{\wt{u}}}}}}$}
\put(60,30){\circle*{5}} \put(64,30){${\large{\wh{\wt{\wt{u}}}}}$}
\put(60,60){\circle*{5}} \put(64,64){${\large{\wt{\wt{u}}}}$}
\put(0,0){\line(0,1){30}} \put(0,0){\line(1,0){30}}
\put(30,30){\line(-1,0){30}} \put(30,30){\line(0,-1){30}}
\put(30,30){\line(1,0){30}} \put(30,30){\line(0,1){30}}
\put(0,60){\line(1,0){30}} \put(0,60){\line(0,-1){30}}
\put(60,0){\line(-1,0){30}} \put(60,0){\line(0,1){30}}
\put(60,60){\line(-1,0){30}} \put(60,60){\line(0,-1){30}}

\end{picture}

\end{center}
\vspace{.6cm}

The notation employed in \eqref{eq:dBSQ} is illustrated in this
Figure: $u=u_{n,m}$ denotes the dependent variable of the lattice
points labelled by $(n,m)\in \mathbb{Z}^2$; $p$, $q$ are continuous
\textit{lattice} parameters associated with the grid size in the
directions of the lattice given by the independent variables $n$ and
$m$ respectively, and for the sake of clarity we prefer to use a
notation with elementary lattice shifts denoted by
\[
u=u_{n,m}~\mapsto~\wt{u}=u_{n+1,m}\quad,\quad
u=u_{n,m}~\mapsto~\wh{u}=u_{n,m+1}\  .
\]
Thus, as a consequence of this notation, we have also
\[ \wh{\wt{u}}=u_{n+1,m+1}\quad.\quad \wh{\wt{\wt{u}}}=u_{n+2,m+1}\quad,\quad \wh{\wh{\wt{u}}}=u_{n+1,m+2}\quad,\quad\wh{\wh{\wt{\wt{u}}}}=u_{n+2,m+2}\  . \]
Together with the lattice BSQ, also lattice versions of the modified
BSQ  (MBSQ) and Schwarzian BSQ equations have been found, cf.
\cite{GD,N1,DIGP,W} as well as their semi-continuum limits (leading
to differential-difference analogues of these equations). The
lattice BSQ equation has (re)gained considerable attention in recent
years, w.r.t. similarity reductions \cite{W,TN2}, soliton solutions
\cite{HZ,MK} and its Lagrangian structure \cite{LN}.

It was shown in \cite{W}, cf. also \cite{TN2}, that eq.
\eqref{eq:dBSQ} can also be written as the following coupled system
of 3-field equations (for fields $u$, $u_{0,1}$ and $u_{1,0}$)
around an elementary quadrilateral: \bse\label{eq:u01}\bea
&& \wt{u}_{0,1}+u_{1,0}=p\wt{u}_0-pu_0+\wt{u}_0 u_0\  , \\
&& \wh{u}_{0,1}+u_{1,0}=q\wh{u}_0-qu_0+\wh{u}_0 u_0\  , \\
&& \frac{p^3-q^3}{p-q+\wh{u}_0-\wt{u}_0} = p^2+pq+q^2
-(p+q)(\wh{\wt{u}}_0-u_0)-u_0\wh{\wt{u}}_0+\wh{\wt{u}}_{1,0}+u_{0,1}\
. \eea\ese In the recent paper \cite{H} a one-parameter
generalization was presented of this system (after a point
transformation on the dependent variables), as well as of two other
3-component BSQ systems, comprising the lattice modified BSQ (MBSQ)
and Schwarzian BSQ (SBSQ) systems, through systematic search based
on the multidimensional consistency property. Subsequently, in
\cite{HZh} soliton solutions for this extended BSQ system were
derived. The main structural feature that emerged from the latter
study was that the dispersion relation defining the relevant
discrete exponential functions involve a cubic of the form
\[ \oa^3-k^3-\beta(\oa^2-k^2)=0\   , \]
where $k$ is like the wave number of the solution, whilst $\oa$ the
corresponding angular frequency. Noting that the parameter $\bb$
corresponds to the deformation of the 3-component system, the limit
$\bb\rightarrow 0$ would lead to the original lattice BSQ system, in
which case the dispersion would involve a cube root of unity for the
parameter $\oa$ (i.e., $\oa\rightarrow \exp(2\pi i/3)k$ in this
limit).

In the present study, we generalize the latter relation to an
arbitrary cubic, i.e. we consider a dispersion of the form
\be\label{eq:disp}
 G(\oa,k):=\oa^3+\alpha_2\oa^2+\alpha_1\oa-\left(k^3+\alpha_2 k^2+\alpha_1 k\right)=0\   , \ee
and show that not only the extended lattice BSQ of \cite{H}, but
also the generalizations of the 3-component systems of lattice MBSQ
and SBSQ, which remained as yet unidentified, arise from one and the
same structure.
We will derive this structure, in the case of the extended
dispersion relation \eqref{eq:disp}, following the direct
linearization scheme developed for the case that
$\alpha_1=\alpha_2=0$ in  \cite{GD}, and subsequently derive from this the main
equations as well as the relations between the various fields arising in
the extended lattice BSQ systems of \cite{H}. In particular we derive the
following coupled system
\bse\label{eq:interol3syst}\bea
&& v_a\wh{\wt{w}}_b= \frac{{\mathcal Q}_{p,q}(u,\wt{u},\wh{u},\wh{\wt{u}})}{Q_aP_b\wt{u}-P_aQ_b\wh{u}}\  , \\
&& \wt{v}_aw_b=P_au-P_b\wt{u}\  , \quad \wh{v}_aw_b=Q_a u-Q_b\wh{u}\   ,
\eea
in which the quadrilateral multilinear function ${\mathcal Q}$ is given by
\be\label{eq:Q}
 {\mathcal Q}_{p,q}(u,\wt{u},\wh{u},\wh{\wt{u}}):= P_a P_b (u \wh{u}+\wt{u}\wh{\wt{u}})
- Q_a Q_b(u\wt{u}+\wh{u}\wh{\wt{u}})+G(-p,-q)  (\wh{u}\wt{u}+ u \wh{\wt{u}})\  ,
\ee\ese
and in which
$$
P_a=\sqrt{-G(-p,-a)} \quad,\quad Q_a=\sqrt{-G(-q,-a)}\quad,\quad
G(p,q)=p^3-q^3+\alpha_2(p^2-q^2)+\alpha_1(p-q)\  .
$$
The system, by elimination of $v_a$ and $w_b$ leads to the following 9-point equation for $u$
\be\label{eq:interBSQ-e}
\frac{{\mathcal Q}_{p,q}(\wh{u},\wh{\wt{u}},\wh{\wh{u}},\wh{\wh{\wt{u}}})}
{{\mathcal Q}_{p,q}(\wt{u},\wt{\wt{u}},\wh{\wt{u}},\wh{\wt{\wt{u}}})}
= \frac{(Q_a u-Q_b\wh{u})\,(P_a \wh{\wh{\wt{u}}}-P_b
\wh{\wh{\wt{\wt{u}}}})\,(Q_aP_b\wh{\wt{u}}-P_a Q_b\wh{\wh{u}})}
{(P_a u-P_b\wt{u})\,(Q_a \wh{\wt{\wt{u}}}-Q_b
\wh{\wh{\wt{\wt{u}}}})\,(Q_aP_b\wt{\wt{u}}-P_a Q_b\wh{\wt{u}})}\ .
\ee
Eq. \eqref{eq:interBSQ-e} is the extended form of the BSQ analogue
of the $({\rm Q3})_0$ equation, and which reduces to the analogous equation
given in \cite{Nij} by setting $\alpha_1=\alpha_2=0$. It is related by a point transformation
to an equation given already in \cite{DIGP,W}, and which constitutes
the rank 3 analogue of the NQC equation of \cite{NQC} (associated
with the lattice KdV equations). In \cite{Nij} the derivation of a
BSQ analogue of the full Q3 equation in the ABS list was presented,
but it gives rise to a rather complicated coupled system. Thus, even though
those results can in principle be readily extended to the more general
dispersion relation \eqref{eq:disp} as well, we will refrain
from doing so in the present paper. As a concrete spin-off of the
derivations presented here, we derive also the Lax pairs for the
extended BSQ system, whilst $N$-soliton solutions in the Cauchy
matrix form can be readily inferred from the direct linearization
structure.

\section{Direct linearization and constitutive system}
\setcounter{equation}{0}

In this section we present the constitutive relations of the direct
linearization scheme for a class of lattice systems generalizing the
lattice Gel'fand-Dikii (GD) hierarchy. The members of this hierarchy
are labelled by an integer $N$, which in the case of \cite{GD} were
associated with the corresponding roots of unity $\oa=\exp(2\pi
i/N)$, the case of $N=2$ corresponding to the lattice KdV systems,
whilst the case $N=3$ yielding the lattice BSQ systems. In the
present paper the root of unity is extended to a dispersion function
$\oa_j(k)$ associated with the roots of arbitrary fixed $N^{\rm th}$
order polynomials.

\subsection{Infinite matrix scheme}

The starting point of the construction is the following integral
\be\label{eq:bC} \bC= \sum_{j=1}^N
\int_{\Gamma_j}\,d\ld_j(k)\,\rho_k
\bc_k\,\bc^t_{-\oa_j(k)}\sg_{-\oa_j(k)}\  , \ee over a yet
unspecified set of contours or arcs $\Gamma_j$ in the complex plane
of a variable $k$, and in which for  $N\geq 2$ a positive integer,
$\oa_j(k)(j=1,2,\cdots,N-1)$ and $\oa_N(k)=k$ are the distinct roots
of the following algebraic relation \be\label{eq:g} G(\oa,k):=
g(\oa)-g(k)=0\quad,\quad {\rm where}\quad  g(k)=\sum_{j=1}^{N}
\alpha_jk^{j}\ \ee is a monic (i.e. $\alpha_N=1$) polynomial with
coefficients $\alpha_j$. The factors $\rho_k$, $\sg_{k'}$ are
discrete exponential functions of $k$, $k'$ respectively, given by
\be\label{eq:rhosgk} \rho_k(n,m)= (p+k)^n
(q+k)^m\rho_k(0,0)\quad,\quad \sg_{k'}(n,m)= (p-k')^{-n}
(q-k')^{-m}\sg_{k'}(0,0)\  , \ee the infinite-component vector
$\bc_k=(k^j)_{j\in\mathbb{Z}}$ denotes the column vector of a basis
of monomials in the variable $k$, and $\bc^t_{k'}$ denotes the
transpose of $\bc_{k'}$. The integration measures $d\ld_j(k)$ remain
unspecified, but we assume that basic operations (such as
differentiations w.r.t. parameters or applying shifts in the
variables $n$ and $m$) commute with the integrations.

In accordance with the notation introduced in section 1, we denote
shifts over one unit in the variables $n$ and $m$ respectively by
the operations $\wt{\phantom{a}}$ and $\wh{\phantom{a}}$, which
implies for the factors $\rho_k$, $\sg_{k'}$
\[ \wt{\rho}_k=\rho_k(n+1,m)=(p+k)\rho_k\quad,\quad  \wh{\rho}_k=\rho_k(n,m+1)=(q+k)\rho_k\ , \]
and
\[ \wt{\sg}_{k'}=\sg_{k'}(n+1,m)=(p-k')^{-1}\sg_{k'}\quad,\quad  \wh{\sg}_{k'}=\sg_{k'}(n,m+1)=(q-k')^{-1}\sg_{k'}\ . \]
These relations imply the following linear relations for the matrix
$\bC$ \be\label{eq:bCdyn} \wt{\bC}\,(p-\tLd)=(p+\Ld)\,\bC\quad,\quad
\wh{\bC}\,(q-\tLd)=(q+\Ld)\,\bC\  , \ee where we have introduced the
matrices $\Ld$ and $\tLd$ which are defined by their actions on the
vector $\bc_k$, and on its transposed vector, as follows:
\be\label{eq:bCninc} \Ld\,\bc_k=k\,\bc_k\quad,\quad
\bc^t_{k'}\tLd=k'\,\bc^t_{k'}\  .
\ee
Besides, from the definition
of $\bC$ together with  \eqref{eq:bCninc}, one has
\bse\label{eq:idens}\be g(\Ld)\,\bC=\bC\,g(-\tLd)\  ,
\label{eq:idens_a} \ee and as a consequence, using the fact that
~$g(\oa)-g(k)=\prod_{j=1}^N(\oa-\oa_j(k))$~, also \be
\left[\prod_{j=1}^{N}(\oa_j(-p)-\Ld)\right]\,\bC=\bC\,\prod_{j=1}^{N}(\oa_j(-p)+\tLd)\
, \label{eq:idens_b} \ee leading to \be
\left[\prod_{j=1}^{N-1}(\oa_j(-p)-\Ld)\right]\,\wt{\bC}=\bC\,\prod_{j=1}^{N-1}(\oa_j(-p)+\tLd)\
, \label{eq:idens_c} \ee\ese (and a similar relation to
\eqref{eq:idens_c} with $p$ replaced by $q$ and $\wt{\bC}$ by
$\wh{\bC}$).
The following ingredients determine the structure:\\
\textit{i)} A Cauchy kernel $\bOm$ defined by the relation
~$\bOm\Ld+\tLd\bOm=\bO$~,
where $\bO$ is a rank 1 projection matrix, obeying $\bO^2=\bO$. \\
\textit{ii)}  A matrix ~$\bU$~ obeying the relation ~$\bU=\bC-\bU\,\bOm\,\bC$.~ \\
\textit{iii)} A vector ~$\bu_k$~ defined by ~$\bu_k=\rho_k(\bc_k-\bU\,\bOm\,\bc_k)$.~\\

\noindent In terms of these objects the following sets of relations
involving the shift $\wt{\phantom{a}}$ can be derived:
\bse\label{eq:NUrels}\bea
\wt{\bU}\,(p-\tLd) &=& (p+\Ld)\bU-\wt{\bU}\,\bO\,\bU\  , \label{eq:NUrels_a} \\
\bU\,\left[ \prod_{j=1}^{N-1}(\oa_j(-p)+\tLd)\right] &=& \left[\prod_{j=1}^{N-1}(\oa_j(-p)-\Ld)\right]\wt{\bU}  \label{eq:NUrels_b}  \\
&& + \bU\, \sum_{j=0}^{N-2} \left[ \prod_{l=1}^j
(\oa_l(-p)+\tLd)\right]\bO \left[\prod_{l=j+2}^{N-1}
(\oa_l(-p)-\Ld)\right]\,\wt{\bU}\   , \nn  \\
 \bU\,\sum_{j=1}^{N}\alpha_j(-\tLd)^{j} &=& \sum_{j=1}^{N}\alpha_j\Ld^{j}\,\bU
-\bU\,\sum_{j=1}^{N} \alpha_j\sum_{l=0}^{j-1}
(-\tLd)^l\bO\Ld^{j-1-l}\,\bU\  . \label{eq:NUrels_c} \eea\ese The
derivation of \eqref{eq:NUrels} follows the same procedure as in
\cite{GD}, and they reduce to the relations obtained there in the
special case that $\oa_j(k)=\exp(2\pi ij/N)k$, i.e. the case that
all coefficients $\alpha_j=0$, $j=1,\dots,N-1$ in \eqref{eq:g}. By
virtue of the covariance of the dynamics in terms of the variables
$n$ and $m$, similar relations to \eqref{eq:NUrels_a} and
\eqref{eq:NUrels_b} hold with the shift $\wt{\phantom{a}}$ replaced
by $\wh{\phantom{a}}$ while replacing $p$ by $q$. The associated
linear problems (Lax pairs) are derived in terms of the object
$\bu_k$, for which we have the following set of constitutive
relation, \cite{GD}, \bse\label{eq:Nukrels}\bea
\wt{\bu}_k &=&  (p+\Ld)\bu_k-\wt{\bU}\,\bO\,\bu_k\  , \label{eq:Nukrels_a} \\
-\left[ \prod_{j=1}^{N}(\oa_j(-p)-k)\right]\bu_k &=& \left[\prod_{j=1}^{N-1}(\oa_j(-p)-\Ld)\right]\wt{\bu}_k \label{eq:Nukrels_b}   \\
&&  +\bU\, \sum_{j=0}^{N-2} \left[ \prod_{l=1}^j
(\oa_l(-p)+\tLd)\right]\bO \left[\prod_{l=j+2}^{N-1}
(\oa_l(-p)-\Ld)\right]\,\wt{\bu}_k \  , \nn
\\
\sum_{j=1}^{N} \alpha_j k^{j}\bu_k &=&  \sum_{j=1}^{N}
\alpha_j\Ld^{j}\,\bu_k -\bU\,\sum_{j=1}^{N} \alpha_j\sum_{l=0}^{j-1}
(-\tLd)^l\bO\Ld^{j-1-l}\,\bu_k\  , \label{eq:Nukrels_c} \eea\ese and
similar relations to \eqref{eq:Nukrels_a} and \eqref{eq:Nukrels_b}
with the shift $\wt{\phantom{a}}$ replaced by $\wh{\phantom{a}}$
while replacing $p$ by $q$.

The relation \eqref{eq:NUrels_b} can be easily derived from the
formal relation
\begin{eqnarray*}
\bU\frac{g(-p)-g(-\tLd)}{-p+\tLd} &=& \frac{g(\Ld)-g(-p)}{p+\Ld}\wt{\bU}  \\
&& + \bU
\left[\bOm\,\frac{g(-p)-g(\Ld)}{p+\Ld}-\frac{g(-p)-g(-\tLd)}{p-\tLd}\,\bOm\right]
\wt{\bU},
\end{eqnarray*}
whilst \eqref{eq:NUrels_c} follows from
\[ \bU g(-\tLd)=g(\Ld)\bU-U(\bOm g(\Ld)-g(-\tLd)\bOm)\bU\  .  \]
Similarly, the linear relation \eqref{eq:Nukrels_b} for $\bu_k$
follows from
\begin{eqnarray*}
G(k,-p) \bu_k &=& \frac{g(\Ld)-g(-p)}{p+\Ld}\wt{\bu}_k  \\
&& + \bU
\left[\bOm\,\frac{g(-p)-g(\Ld)}{p+\Ld}-\frac{g(-p)-g(-\tLd)}{p-\tLd}\,\bOm\right]
\wt{\bu}_k ,
\end{eqnarray*}
whereas \eqref{eq:Nukrels_c} follows from
\[  g(k) \bu_k=g(\Ld)\bu_k-\bU(\bOm g(\Ld)-g(-\tLd)\bOm)\bu_k\  . \]

For any fixed value of $N$, these abstract equations form an
infinite set of recurrence relations defining the dynamics in terms
of the independent variables $n$, $m$ on the objects $\bU$ and
$\buk$ taking values in an abstract vector space $\mathcal{V}$, and
an adjoint vector $\tbme$ in its dual $\mathcal{V}^{\ast}$. To give
a concrete realization, we can choose a fixed vector $\bme$ and use
the matrices $\Ld$ and $\tLd$ to define a natural gradation in this
vector space, In terms these, we realize quantities such as $\bC$
and $\bU$ as infinite-dimensional matrices. In fact, setting
\be\label{eq:uijdef}
 u_{i,j}:= \tbme\Ld^i\bU\tLd^j\bme\quad,\quad i,j\in\mathbb{Z}\   ,
\ee we obtain infinite, i.e., $\mathbb{Z}\times\mathbb{Z}$ matrices,
on which $\Ld$ and $\tLd$ act as index-raising operators acting from
the left and from the right. In this realization the matrix
~$\bO=\bme\,\tbme$~ is the projector defined by these vectors, and
through the relations 
\be\label{eq:cte}
\tbme\,\bc_k=1\  , \quad \bc^t_{k'}\,\bme=1\  ,
\ee 
this defines the ``central'' entry in the space of
$\mathbb{Z}\times\mathbb{Z}$ matrices. We will define concrete
objects in terms of the matrix $\bU$, where we allude to this
realization.

\section{Lattice BSQ case}

\setcounter{equation}{0}

In this section we will now apply the structure exhibited in section
2 to the case $N=3$, which in \cite{GD} is the one leading to the
BSQ class of systems. We recall that the case $N=2$ leads to the KdV
class of lattice systems, but in that case it can be shown that the
extended dispersion relation does not lead to new results. However,
in the BSQ case ($N=3$) it does. We restrict ourselves in this paper
primarily to this case of $N=3$, but it is clear that the general
system derived in section 2 can be studied in fairly similar ways
for any (integer) $N>2$.  Most of the relations given in
this section reduce to the ones given in \cite{W} setting
$\alpha_1=\alpha_2=0$.

\subsection{Basic objects and their relations: BSQ case $(N=3)$}

Let us define \be\label{eq:P(oa-k)}
G(\oa,k):=\oa^3-k^3+\alpha_2(\oa^2-k^2)+\alpha_1(\oa- k). \ee
Obviously, for $N=3$ the algebraic relation \eqref{eq:g} reduces to
\be\label{eq:bC-N3} G(\oa,k)=0, \ee and its roots are denoted by
$\oa_1(k),~\oa_2(k)$ and $\oa_3(k)=k$.
The set of relations \eqref{eq:NUrels} take the form:
\bse\label{eq:Urels}\bea
\wt{\bU}\,(p-\tLd)&=& (p+\Ld)\bU-\wt{\bU}\,\bO\,\bU\  , \label{eq:Urels_a} \\
\bU\,(\oa_1(-p)+\tLd)(\oa_2(-p)+\tLd) &=& (\oa_1(-p)-\Ld)(\oa_2(-p)-\Ld)\wt{\bU} \nn \\
&&~~+\bU\,\left[ (p-\alpha_2)\bO-(\bO\,\Ld-\tLd\,\bO)\right]\,\wt{\bU} \  , \label{eq:Urels_b} \\
\bU\,\sum_{j=1}^{3}\alpha_j(-\tLd)^{j} &=&
\sum_{j=1}^{3}\alpha_j\Ld^{j}\,\bU -\bU\,\sum_{j=1}^{3}
\alpha_j\sum_{l=0}^{j-1} (-\tLd)^l\bO\Ld^{j-1-l}\,\bU\   ,
\label{eq:Urels_c} \eea\ese and similar relations to
\eqref{eq:Urels_a} and \eqref{eq:Urels_b} with the shift
$\wt{\phantom{a}}$ replaced by $\wh{\phantom{a}}$ while replacing
$p$ by $q$, as well as \bse\label{eq:ukrels}\bea
\wt{\bu}_k &=& (p+\Ld)\bu_k-\wt{\bU}\,\bO\,\bu_k\  , \label{eq:ukrels_a} \\
-\left[ \prod_{j=1}^{3}(\oa_j(-p)-k)\right]\bu_k &=& (\oa_1(-p)-\Ld)(\oa_2(-p)-\Ld)\wt{\bu}_k \nn \\
&&~~+\bU\,\left[ (p-\alpha_2)\bO-(\bO\,\Ld-\tLd\,\bO)\right]\,\wt{\bu}_k \  , \label{eq:ukrels_b} \\
\sum_{j=1}^{3} \alpha_j k^{j}\bu_k &=& \sum_{j=1}^{3}
\alpha_j\Ld^{j}\,\bu_k -\bU\,\sum_{j=1}^{3} \alpha_j\sum_{l=0}^{j-1}
(-\tLd)^l\bO\Ld^{j-1-l}\,\bu_k\  , \label{eq:ukrels_c} \eea\ese (and
similar relations to \eqref{eq:ukrels_a} and \eqref{eq:ukrels_b}
with the shift $\wt{\phantom{a}}$ replaced by $\wh{\phantom{a}}$
while replacing $p$ by $q$). For the sake of obtaining from the
constitutive relations \eqref{eq:Urels} closed-form equations, we
introduce the following objects:
\bse\label{eq:objs}\bea
&& v_a: = 1-\tbme\,(a+\Ld)^{-1}\,\bU\,\bme\quad,\quad w_b:= 1+\tbme\,\bU\,(b-\tLd)^{-1}\,\bme\  , \\
&& s_a: = a-\tbme\,(a+\Ld)^{-1}\,\bU\,\tLd\,\bme\quad,\quad t_b:= -b+\tbme\,\Ld\,\bU\,(b-\tLd)^{-1}\,\bme\  , \\
&& r_a: = a^2-\tbme\,(a+\Ld)^{-1}\,\bU\,\tLd^2\,\bme\quad,\quad
z_b:= b^2+\tbme\,\Ld^2\,\bU\,(b-\tLd)^{-1}\,\bme\  , \eea\ese
but in particular the following object:
\begin{equation}\label{eq:s}
 s_{a,b}:= \tbme\,(a+\Ld)^{-1}\,\bU\,(-b+\tLd)^{-1}\bme.
\end{equation}
Then for these objects and ${u}_{i,j}$ defined in \eqref{eq:uijdef}
we have the following relations \bse\label{eq:uij}\bea
&& p\wt{u}_{i,j}-\wt{u}_{i,j+1}=pu_{i,j}+u_{i+1,j}-\wt{u}_{i,0} u_{0,j}\  , \label{eq:uij_a} \\
&& (p^2-\alpha_2p+\alpha_1)u_{i,j}+(p-\alpha_2)u_{i,j+1}+u_{i,j+2}\nonumber\\
&&=(p^2-\alpha_2p+\alpha_1)\wt{u}_{i,j}-(p-\alpha_2)\wt{u}_{i+1,j}+\wt{u}_{i+2,j}
   +(p-\alpha_2)u_{i,0}\wt{u}_{0,j}-u_{i,0}\wt{u}_{1,j}+u_{i,1}\wt{u}_{0,j}\  , \nn \\
   \label{eq:uij_b}
\eea\ese as well as \bse\label{eq:ssrels}\bea
&& 1+(p-a)s_{a,b}-(p-b)\wt{s}_{a,b}=\wt{v}_a w_b\  , \label{eq:ssrels_a} \\
&& (p+a+b-\alpha_2)+p_bs_{a,b}-p_a\wt{s}_{a,b} = s_a \wt{w}_b-v_a
\wt{t}_b+ (p-\alpha_2)v_a \wt{w}_b\  , \label{eq:ssrels_b} \eea\ese
and \bse\bea \label{eq:strels}
\wt{s}_a &=& (p+u_0)\wt{v}_a-(p-a) v_a \  , \label{eq:strels_a} \\
t_b &=& (p-b)\wt{w}_b-(p-\wt{u}_0)w_b\  , \label{eq:strels_b}
\eea\ese and \bse\label{eq:zrrels}\bea
\wt{r}_a &=& p\wt{s}_a-(p-a) s_a+\wt{v}_a u_{0,1}\  , \label{eq:zrrels_a}\\
z_b &=& (p-b) \wt{t}_b-p t_b+\wt{u}_{1,0} w_b\  ,
\label{eq:zrrels_b} \eea\ese and \bse\label{eq:rzrels}\bea
r_a &=& p_a\wt{v}_a-(p-\wt{u}_0-\alpha_2)s_a-\left((p-\alpha_2)(p-\wt{u}_0)+\wt{u}_{1,0}+\alpha_1 \right) v_a\  , \label{eq:rzrels_a} \\
\wt{z}_b &=& p_b w_b+(p+u_0-\alpha_2)\wt{t}_b-\left(
(p-\alpha_2)(p+u_0)+u_{0,1}+\alpha_1\right) \wt{w}_b\  ,
\label{eq:rzrels_b} \eea\ese where we have used a simplified
notation $u_0=u_{0,0}$ and $p_a$, $p_b$ are defined as
\be\label{eq:papb}
p_a=\frac{G(-p,-a)}{a-p}=(p^2+ap+a^2)-\alpha_2(p+a)+\alpha_1\  ,\quad
p_b=\frac{G(-p,-b)}{b-p}=(p^2+bp+b^2)-\alpha_2(p+b)+\alpha_1\  . \ee
All relations \eqref{eq:ssrels}-\eqref{eq:rzrels}, which are in a
slightly different form have already appeared in
\cite{W}\footnote{We note in passing that the yet unidentified case
(A) of 3-component BSQ type systems which was presented in the
recent paper \cite{H} can be identified with eqs. (5.3.7a) together
with (5.3.14) of \cite{W}, where, up to a point transformation. $x$
can be identified with $v_\alpha$, $z$ with $u$ and $y$ with
$s_\alpha$. By duality this case of \cite{H} can also be identified
with eqs. (5.3.7b) and (5.3.16) of \cite{W} identifying $x$ with
$w_\beta$, $z$ with $u$ and $y$ with $t_\beta$.}, also hold for
their hat-$q$ counterparts obtained by replacing the shift
$\wt{\phantom{a}}$ by the shift $\wh{\phantom{a}}$ whilst replacing
the parameter $p$ by $q$.

From equations \eqref{eq:uij_a}, \eqref{eq:strels_a},
\eqref{eq:rzrels_a} and their hat-$q$ counterparts one can get a
further set of relations \bse\label{eq:svurels}\bea
&& (p+q-\wh{\wt{u}}_0+\frac{s_a}{v_a}-\alpha_2)(p-q+\wh{u}_0-\wt{u}_0)=p_a\frac{\wt{v}_a}{v_a}-q_a\frac{\wh{v}_a}{v_a}\  , \label{eq:svurels_a} \\
&&
p-q+\wh{u}_0-\wt{u}_0=(p-a)\frac{\wh{v}_a}{\wh{\wt{v}}_a}-(q-a)\frac{\wt{v}_a}{\wh{\wt{v}}_a}\
, \label{eq:svurels_b} \eea\ese and similarly, from
\eqref{eq:uij_a}, \eqref{eq:strels_b} and \eqref{eq:rzrels_b},
\bse\label{eq:twurels}\bea &&
(p+q+u_0-\frac{\wh{\wt{t}}_b}{\wh{\wt{w}}_b}-\alpha_2)(p-q+\wh{u}_0-\wt{u}_0)=p_b\frac{\wh{w}_b}{\wh{\wt{w}}_b}-q_b\frac{\wt{w}_b}{\wh{\wt{w}}_b}\
,
\label{eq:twurels_a} \\
&&
p-q+\wh{u}_0-\wt{u}_0=(p-b)\frac{\wt{w}_b}{w_b}-(q-b)\frac{\wh{w}_b}{w_b}\
. \label{eq:twurels_b} \eea\ese

\subsection{Closed-form lattice equations}

One can immediately get two closed-form lattice equations. One is
composed of \eqref{eq:strels_a}, its hat-$q$ counterpart and
\eqref{eq:svurels_a}, i.e.,
\bse\label{eq:bA2}\bea
&& \wt{s}_a = (p+u_0)\wt{v}_a-(p-a) v_a \ , ~~~~\wh{s}_a = (q+u_0)\wh{v}_a-(q-a) v_a \ ,\label{eq:bA2-a}\\
&&
(p+q-\wh{\wt{u}}_0+\frac{s_a}{v_a}-\alpha_2)(p-q+\wh{u}_0-\wt{u}_0)=p_a\frac{\wt{v}_a}{v_a}-q_a\frac{\wh{v}_a}{v_a}\
. \label{eq:bA2-b}
\eea\ese
Another is composed of
\eqref{eq:strels_b}, its hat-$q$ counterpart and
\eqref{eq:twurels_b}, i.e., \bse\label{eq:bA2-c}\bea
&& t_b =(p-b)\wt{w}_b-(p-\wt{u}_0)w_b\  , ~~~~t_b= (q-b)\wh{w}_b-(q-\wh{u}_0)w_b\  , \label{eq:bA2-c-a}\\
&&
(p+q+u_0-\frac{\wh{\wt{t}}_b}{\wh{\wt{w}}_b}-\alpha_2)(p-q+\wh{u}_0-\wt{u}_0)=p_b\frac{\wh{w}_b}{\wh{\wt{w}}_b}-q_b\frac{\wt{w}_b}{\wh{\wt{w}}_b}\
. \label{eq:bA2-c-b} \eea\ese

To get further closed form lattice systems we take $i=j=0$ in
\eqref{eq:uij} and its hat-$q$ counterpart and get
\bse\label{eq:u00}\bea &&
p\wt{u}_{0}-\wt{u}_{0,1}=pu_{0}+u_{1,0}-\wt{u}_{0} u_{0}\  , ~~~~~~~
 q\wh{u}_{0}-\wh{u}_{0,1}=qu_{0}+u_{1,0}-\wh{u}_{0} u_{0}\  , \label{eq:u00_ab} \\
&& (p^2-\alpha_2p+\alpha_1)u_{0}+(p-\alpha_2)u_{0,1}+u_{0,2}\nonumber\\
&&=(p^2-\alpha_2p+\alpha_1)\wt{u}_{0}-(p-\alpha_2)\wt{u}_{1,0}+\wt{u}_{2,0}
   +(p-\alpha_2)u_{0}\wt{u}_{0}-u_{0}\wt{u}_{1,0}+u_{0,1}\wt{u}_{0}\  \label{eq:u00_c} , \\
&& (q^2-\alpha_2q+\alpha_1)u_{0}+(q-\alpha_2)u_{0,1}+u_{0,2}\nonumber\\
&&=(q^2-\alpha_2q+\alpha_1)\wh{u}_{0}-(q-\alpha_2)\wh{u}_{1,0}+\wh{u}_{2,0}
   +(q-\alpha_2)u_{0}\wh{u}_{0}-u_{0}\wh{u}_{1,0}+u_{0,1}\wh{u}_{0}\ . \label{eq:u00_d}
\eea\ese Let us focus on \eqref{eq:u00_c} and \eqref{eq:u00_d}. By
substraction one can delete $u_{0,2}$, and to delete $u_{2,0}$ from
the remains one can make use of \eqref{eq:uij_a} where we take
$i=1,~j=0$. Then, after some algebra we can reach to
\be\label{eq:BSQ-eq3}
\frac{-G(-p,-q)}{p-q+\wh{u}_0-\wt{u}_0}=\frac{G(-p,-q)}{q-p}
+(p+q-\alpha_2)(u_0- \wh{\wt{u}}_0)-u_0
\wh{\wt{u}}_0+\wh{\wt{u}}_{1,0}+u_{0,1}. \ee This equation together
with \eqref{eq:u00_ab} composes of a closed-form system which is
viewed as the extended three-component lattice BSQ equation. (We
make a precise connection with the corresponding system of \cite{H}
in the next section).

Next, by the transformation (if $a\neq b$)
\be\label{eq:Sab-sab}
s_{a,b}=S_{a,b}-\frac{1}{b-a} \ee we rewrite the relation
\eqref{eq:ssrels} and its hat-$q$ counterpart in the form
\bse\label{eq:SSrels}\bea && (p-a)S_{a,b}-(p-b)\wt{S}_{a,b}=\wt{v}_a
w_b\  ,
~~~~~~(q-a)S_{a,b}-(q-b)\wh{S}_{a,b}=\wh{v}_a w_b\ , \label{eq:SSrels_ab} \\
&& p_bS_{a,b}-p_a\wt{S}_{a,b} = s_a \wt{w}_b-v_a \wt{t}_b+
(p-\alpha_2)v_a \wt{w}_b\ ,
\label{eq:SSrels_c}\\
&& q_bS_{a,b}-q_a\wh{S}_{a,b} = s_a \wh{w}_b-v_a \wh{t}_b+
(q-\alpha_2)v_a \wh{w}_b\  . \label{eq:SSrels_d}
\eea\ese
Then, from
\eqref{eq:SSrels_c}$\wh{\phantom{a}}$ deleting $\wh{\wt{t}}_{b}$ by
using \eqref{eq:twurels_a} and $\wh{s}_a$ by using the hat-$q$
version of \eqref{eq:strels_a} and also making use of
\eqref{eq:twurels_b} we get
\be\label{eq:vwSrels}
(q-a)v_{a}\wh{\wt{w}}_{b}=\frac{p_b\wh{w}_b-q_b\wt{w}_b}
{(p-b)\wt{w}_b-(q-b)\wh{w}_b}\wh{v}_{a}w_b+p_a\wh{\wt{S}}_{a,b}-p_b\wh{S}_{a,b}\ .
\ee
Using \eqref{eq:SSrels_ab} once again we can rewrite
\eqref{eq:vwSrels} to the following  desired form
\be\label{eq:vwSrels-1}
v_{a}\wh{\wt{w}}_{b}=w_b\frac{\frac{p_b}{p-a}\wh{w}_b\wt{v}_a-\frac{q_b}{q-a}\wt{w}_b\wh{v}_a}{(p-b)\wt{w}_b-(q-b)\wh{w}_b}
+\frac{G(-a,-b)}{(p-a)(q-a)}\wh{\wt{S}}_{a,b}\ .
\ee
The system composed of \eqref{eq:vwSrels-1} and \eqref{eq:SSrels_ab} can be
viewed as the extended three-component lattice MBSQ/SBSQ system, in
the sense that the extended versions of both the 9-point lattice
MBSQ of \cite{GD} as well as of the 9-point Schwarzian BSQ lattice
can be obtained from this system by elimination of two of the three
variables. An (equivalent) alternative form of the  extended
three-component lattice MBSQ/SBSQ equation consists of
\eqref{eq:SSrels_ab} together with the following equation
\be\label{eq:vwSrels-2}
v_{a}\wh{\wt{w}}_{b}=w_b\frac{\frac{p_a}{p-b}\wh{w}_b\wt{v}_a-\frac{q_a}{q-b}\wt{w}_b\wh{v}_a}
{(p-b)\wt{w}_b-(q-b)\wh{w}_b}
+\frac{G(-a,-b)}{(p-b)(q-b)}S_{a,b}\ . \ee \eqref{eq:vwSrels-2} can
be derived similar to \eqref{eq:vwSrels-1}. We note that
\eqref{eq:vwSrels-1} and \eqref{eq:vwSrels-2} can be transformed to
each other in the view of \eqref{eq:SSrels_ab}. In fact, noting that
the relation $G(-p,-b)-G(-p,-a)=G(-a,-b)$, a substraction of
\eqref{eq:vwSrels-2} from \eqref{eq:vwSrels-1} gives
\[\frac{\wh{\wt{S}}_{a,b}}{(p-a)(q-a)}-\frac{S_{a,b}}{(p-b)(q-b)}
=\frac{\frac{\wh{w}_b\wt{v}_a
w_b}{(p-a)(p-b)}-\frac{\wt{w}_b\wh{v}_a
w_b}{(q-a)(q-b)}}{(p-b)\wt{w}_b-(q-b)\wh{w}_b} \ ,
\]
which, by replacing $\wt{v}_a w_b$ and $\wh{v}_a w_b$ using
\eqref{eq:SSrels_ab}, further reduces to
\[ \wh{\wt{S}}_{a,b}
=\frac{(p-a) \wt{w}_b\wh{S}_{a,b} -(q-a) \wh{w}_b\wt{S}_{a,b}
}{(p-b)\wt{w}_b-(q-b)\wh{w}_b} \ ,
\]
and this is an identity as is evident from \eqref{eq:SSrels_ab}.

Finally, to derive the coupled system \eqref{eq:interol3syst} we rewrite \eqref{eq:vwSrels} in the form
\bea\label{eq:vwSrels-r}
 (q-a)v_{a}\wh{\wt{w}}_{b} &=&
 \frac{p_b\wh{\wt{v}}_a\wh{w}_b-q_b\wh{\wt{v}}_a\wt{w}_b}{(p-b)\wh{\wt{v}}_a\wt{w}_b-(q-b)\wh{\wt{v}}_a\wh{w}_b}\wh{v}_{a}w_b
+p_a\wh{\wt{S}}_{a,b}-p_b\wh{S}_{a,b} \nn \\
\qquad  &=&
\frac{p_b[(p-a)\wh{S}_{a,b}-(p-b)\wh{\wt{S}}_{a,b}]-q_b[(q-a)\wt{S}_{a,b}-(q-b)\wh{\wt{S}}_{a,b}]}
{(p-b)(q-a)\wt{S}_{a,b}-(q-b)(p-a)\wh{S}_{a,b}} \nn \\
&&\times[(q-a)S_{a,b}-(q-b)\wh{S}_{a,b}]
+p_a\wh{\wt{S}}_{a,b}-p_b\wh{S}_{a,b}\ ,
\eea
and introduce the variable $u$ by setting
\be\label{eq:Sab-u}
S_{a,b}=\bigg(\frac{(p-a)P_b}{(p-b)P_a}\bigg)^n
\bigg(\frac{(q-a)Q_b}{(q-b)Q_a}\bigg)^m u \ . \ee
Furthermore, \eqref{eq:interBSQ-e} follows from \eqref{eq:interol3syst} by
making use of the equality
\be\label{eq:vweq}
 \frac{(v_a\wh{\wt{w}}_b)^{\wh{}}}{(v_a\wh{\wt{w}}_b)^{\wt{}}}
 =\frac{\wh{v}_aw_b}{\wt{v}_a w_b}\cdot \frac{\wh{\wh{\wt{\wt{v}}}}_a\wh{\wh{\wt{w}}}_b}{\wh{\wh{\wt{\wt{v}}}}_a \wh{\wt{\wt{w}}}_b} \ .
\ee
We have now all the ingredients in place to make the identifications with the systems exhibited in \cite{H}.

\section{Deformation of the extended lattice BSQ-type equations}

In this section, we identify the extended closed-form lattice BSQ
equations obtained in the previous section, with the four extended
lattice BSQ-type equations (A-2), (B-2), (C-3) and (C-4) of ref.
\cite{H} by means of simple point transformations. \vspace{.3cm}

\noindent
{\bf \underline{Case A-2}:} \\
Starting from \eqref{eq:bA2} and introducing point transformation
\be\label{eq:xyz-A2} v_a = \frac{x}{x_{a}}  , ~~ u_0=z-z_0,~~
s_a=\frac{1}{x_a}(y-v_ay_a)\ , \ee where \bse\label{eq:xyz-0-A2}\bea
x_a &=& (p-a)^{-n}(q-a)^{-m}c_1\  , \\
z_0 &=& (c_3-p)n+(c_3-q)m+c_2\  , \\
y_a &=& x_a(z_0-c_3)\ , \eea\ese and $c_1$, $c_2$, $c_3$ are
constants, \eqref{eq:bA2} yields \bse\label{eq:A2-ex} \bea
&& \wt{y} = z\wt{x}-x,~~\wh{y} = z\wh{x}-x\  , \\
&& y =x\wh{\wt{z}}-b_0
x+\frac{-G(-p,-a)\,\wt{x}+G(-q,-a)\,\wh{x}}{\wh{z}-\wt{z}}\
,\label{eq:4.3b} \eea\ese where $b_0=-\alpha_2+3c_3$. This is the
(A-2) equation given in ref. \cite{H} and the parameter $b_0$ can be
removed by a transformation \cite{H}.

Alternatively, employing the point transformation
\be\label{eq:xyz-A2-1} w_b = \frac{x}{x_{b}}  , ~~ u_0=z-z_0,~~
t_b=\frac{1}{x_b}(y-w_by_b)\ , \ee where
\bse\label{eq:xyz-0-A2-1}\bea
x_b &=& (-p+b)^{n}(-q+b)^{m}c_1\  , \\
z_0 &=& -(c_3+p)n-(c_3+q)m+c_2\  , \\
y_b &=& x_b(z_0-c_3)\ , \eea\ese and $c_1$, $c_2$, $c_3$ are
constants, we find \bse\label{eq:A2-ex-1}\bea
&& y =\wt{z} x-\wt{x},~~ y = \wh{z}x-\wh{x} \  , \\
&& \wh{\wt{y}} = \wh{\wt{x}}z-b'_0
\wh{\wt{x}}+\frac{-G(-p,-b)\wh{x}+G(-q,-b)\wt{x}}{\wh{z}-\wt{z}}\ ,
\eea\ese where $b'_0=\alpha_2+3c_3$. This equation is related to
\eqref{eq:A2-ex} by the transformation
\[p\rightarrow -p,~q\rightarrow -q, ~n\rightarrow -n, ~m\rightarrow -m,~ \alpha_2\rightarrow -\alpha_2,~b \rightarrow -a, \]
which corresponds to the reversal symmetry of (A-2), (see \cite{H}).

\vspace{.3cm}

\noindent
{\bf \underline{Case B-2}:} \\
Now we look at the extended three-component lattice BSQ equation
\eqref{eq:u00_ab} and \eqref{eq:BSQ-eq3}. Making transformation
\bse\label{eq:xyz-ex}\bea
x &=& u_0-x_{0}\  , \\
y &=& u_{1,0}-x_{0}u_0+y_0\  , \\
z &=& u_{0,1}-x_{0}u_0+z_0\ , \eea\ese with \bse\label{eq:xyz-0}\bea
x_0 &=& np+mq+c_1\  , \\
y_0 &=& \frac{1}{2}(np+mq+c_1)^2+\frac{1}{2}(np^2+mq^2+c_2)+c_3\  , \\
z_0 &=& \frac{1}{2}(np+mq+c_1)^2-\frac{1}{2}(np^2+mq^2+c_2)-c_3\ ,
\eea\ese and constants $c_j, (j=1,2,3)$, we have
\bse\label{eq:BSQ-ex}\bea
&& \wt{z} = x\wt{x}-y,~~\wh{z} = x\wh{x}-y\  , \\
&& z =
x\wh{\wt{x}}-\wh{\wt{y}}-\alpha_2(\wh{\wt{x}}-x)-\alpha_1-\frac{G(-p,-q)}{\wh{x}-\wt{x}}\
. \label{eq:4.9b} \eea\ese This is the (B-2) in Ref.\cite{H} and the
parameter $\alpha_1$ can be removed by transformation \cite{H}.
\vspace{.3cm}

\noindent
{\bf \underline{Case C-3}:} \\
Next, we come to the extended three-component lattice MBSQ/SBSQ
equation \eqref{eq:SSrels_ab} and \eqref{eq:vwSrels-1}. By the
transformation \bse\label{eq:MSBSQ-tran}\bea
S_{a,b} &=& \bigg(\frac{p-a}{p-b}\bigg)^n\bigg(\frac{q-a}{q-b}\bigg)^m x\  , \\
v_a &=& (p-a)^n(q-a)^m y\  , \\
w_b &=& (p-b)^{-n}(q-b)^{-m} z\ , \eea\ese then we have
\bse\label{eq:xyz-MSBSQ}\bea x-\wt{x} &=& \wt{y}z\  , ~~
x-\wh{x} = \wh{y}z\  , \label{eq:xyz-MSBSQ-a} \\
y \wh{\wt{z}} &=&
z\frac{-G(-p,-b)\wh{z}\wt{y}+G(-q,-b)\wt{z}\wh{y}}{\wt{z}-\wh{z}}+G(-a,-b)\wh{\wt{x}}\
. \label{eq:xyz-MSBSQ-b} \eea\ese Similarly, by the same
transformation from \eqref{eq:SSrels_ab} and \eqref{eq:vwSrels-2} we
have \bse\label{eq:xyz-MSBSQ-1}\bea x-\wt{x} &=& \wt{y}z\  , ~~
x-\wh{x} = \wh{y}z\ , \label{eq:xyz-MSBSQ-1a} \\
y \wh{\wt{z}} &=&
z\frac{-G(-p,-a)\wh{z}\wt{y}+G(-q,-a)\wt{z}\wh{y}}{\wt{z}-\wh{z}}+G(-a,-b)x\
. \label{eq:xyz-MSBSQ-1b} \eea\ese This gives (C-3) equation in
\cite{H}. \eqref{eq:xyz-MSBSQ-1} and \eqref{eq:xyz-MSBSQ} are
related by
\[n\rightarrow -n,~m\rightarrow -m,~y\rightarrow z,~z\rightarrow -y,~a \leftrightarrow b\]
which gives the reversal symmetry of (C-3)\cite{H}. Besides, as we
mentioned before,  \eqref{eq:xyz-MSBSQ-b} and
\eqref{eq:xyz-MSBSQ-1b} are also connected through
$$ z\frac{\wh{z}\wt{y}-\wt{z}\wh{y}}{\wt{z}-\wh{z}}=\wh{\wt{x}}-x,$$
which holds in light of \eqref{eq:xyz-MSBSQ-a}. \vspace{.3cm}

\noindent
{\bf \underline{Case C-4}:} \\
The identification of Case (C-4) of \cite{H} is somewhat indirect
and relies on the important observation that \eqref{eq:xyz-MSBSQ}
and \eqref{eq:xyz-MSBSQ-1} share the same solution
\eqref{eq:MSBSQ-tran}. Thus, from these realtions we have
\bse \label{eq:C3-s} \bea
x - \wt{x} &=& \wt{y}z\  , ~~ x-\wh{x}= \wh{y}z , \label{eq:C3-s1} \\
y \wh{\wt{z}} &=&
z\frac{P_{a,b}\,\wh{z}\wt{y}-Q_{a,b}\,\wt{z}\wh{y}}{\wt{z}-\wh{z}}+G_{a,b}(\wh{\wt{x}}+x)
\ , \label{eq:C3-s2} \eea\ese where \bse
\begin{align}
P_{a,b}&=-\frac{1}{2}(G(-p,-b)+G(-p,-a)),\\
Q_{a,b}& =-\frac{1}{2}(G(-q,-b)+G(-q,-a)),\\
G_{a,b}&= \frac{G(-a,-b)}{2}.
\end{align}
\ese

Consider the following transformation \be
x=\frac{x_1-G_{a,b}}{2G_{a,b}(x_1+G_{a,b})},~~
y=\frac{y_1}{x_1+G_{a,b}},~~ z=\frac{z_1}{x_1+G_{a,b}}.
\label{eq:C3-C4} \ee Imposing this transformation on
\eqref{eq:C3-s}, \eqref{eq:C3-s1} yields \be x_1-\wt{x}_1=
\wt{y}_1z_1\  , ~~ x_1-\wh{x}_1= \wh{y}_1z_1, \label{eq:C4-a} \ee
which further gives \be
\wh{\wt{x}}_1=\frac{\wh{x}_1\wt{z}_1-\wt{x}_1\wh{z}_1}{\wt{z}_1-\wh{z}_1}\
. \label{eq:con-a} \ee Meanwhile, \eqref{eq:C3-s2} multiplied by
$(G_{a,b}+x_1)(G_{a,b}+\wh{\wt x}_1)$ yields \be y_1 \wh{\wt
z}_1=\Delta+ x_1 \wh{\wt x}_1-G^2_{a,b}, \ee where \be
\Delta=\frac{(G_{a,b}+\wh{\wt
x}_1)z_1(P_{a,b}\,\wh{z}_1\wt{y}_1-Q_{a,b}\,\wt{z}_1\wh{y}_1)}
{(\wh{x}_1\wt{z}_1-\wt{x}_1\wh{z}_1)+G_{a,b}(\wt{z}_1-\wh{z}_1)}.
\ee Substituting \eqref{eq:con-a} in to $\Delta$ we then have \be
y_1 \wh{\wt
z}_1=z_1\,\frac{P_{a,b}\,\wh{z}_1\wt{y}_1-Q_{a,b}\,\wt{z}_1\wh{y}_1}{\wt{z}_1-\wh{z}_1}+
x_1 \wh{\wt x}_1-G^2_{a,b}. \label{eq:C4-b} \ee \eqref{eq:C4-a} and
\eqref{eq:C4-b} constitute the (C-4) equation of \cite{H}.
\vspace{.3cm}

In conclusion, we have in this section identified all deformed
equations of extended lattice BSQ type of \cite{H} with equations in
section 3 arising there from the direct linearization scheme of
section 2. We note that, whilst the results in \cite{H} were
obtained from a systematic search programme, the equations that are
found from the classification conditions come somewhat as separate
cases through the elimination of choices and the implementation of
certain symmetries. However, search does not really reveal the
interconnections between the variables of the different cases, and,
in contrast, that is what the DL approach provides. Thus, deriving
all equations from the single framework all the interconnections
between the variables of the various systems are revealed. In the next section
we show how explicit soliton solutions are obtained systematically from the DL scheme outlined
in section 2 and applied to the choices of variables associated with the
extended BSQ systems.

\section{Soliton solutions}

\setcounter{equation}{0}

In the case of soliton solutions, we start from the following explicit expression for the infinite
matrix $\bC$
\begin{equation}\label{eq:Csoliton-1}
\bC= \sum_{j=1}^3 \sum_{j'=1}^{N_j} \Lambda_{j,j'}
\rho_{k_{j,j'}}\,
\bc_{k_{j,j'}}\,\tbc_{-\oa_j(k_{j,j'})}\sg_{-\oa_j(k_{j,j'})}\ ,
\end{equation}
which can be obtained by choosing in \eqref{eq:bC} the integration contours $\Gamma_j$ in the
complex plane of the variable $k$ to be Jordan curves  surrounding simple poles at locations
$k_{j,j'}$ in the complex plane, with residues $\Lambda_{j,j'}$
of the measures $d\lambda_j(k)$, assuming the latter to be meromorphic in $k$.
In this case we can derive the following expression for the infinite matrix $\bU$
\begin{equation}\label{eq:Usoliton-1}
\bU= \sum_{j=1}^3 \sum_{j'=1}^{N_j} \Lambda_{j,j'}
\bu_{k_{j,j'}}\,\tbc_{-\oa_j(k_{j,j'})} \sg_{-\oa_j(k_{j,j'})} \ .
\end{equation}
in which the infinite-component vector $\bu_k$ obeys the linear equation:
\begin{equation}\label{eq:uksoliton-1}
\bu_k +\sum_{j=1}^3 \sum_{j'=1}^{N_j} \Lambda_{j,j'}\bu_{k_{j,j'}}\,\frac{\rho_k\,\sg_{-\oa_j(k_{j,j'})}}{k-\oa_j(k_{j,j'})}=\rho_k \bc_k \ .
\end{equation}
Eqs. \eqref{eq:Usoliton-1} and \eqref{eq:uksoliton-1} are derived from the basic definitions in the Appendix.

By setting in eq. \eqref{eq:uksoliton-1} $k=k_{j',i'}$, where $j'=1,2,3$, $i'=1,\dots, N_{j'}$, the equation becomes a linear
system for the quantities $u_{k_{j,j'}}$, with a finite block-Cauchy matrix of the form:
\begin{equation}\label{eq:MCauchy}
\bM_{I,J}=\left( M_{(i,i'),(j,j')}\right)_{i,j=1,2,3;i'=1,\dots,N_i,j'=1,\dots,N_j}:=
\frac{\rho_{k_{i,i'}}\Lambda_{j,j'}\sg_{-\oa_j(k_{j,j'})}}{k_{i,i'}-\oa_j(k_{j,j'})}\  ,
\end{equation}
which is a 3$\times$3 block matrix with rectangular blocks of size $N_i\times N_j$. We use the capital compound indices
$I=(i,i')$ and $J=(j,j')$ to simplify the notation.

We can make the soliton solutions now explicit by assuming that the coefficients $\Ld_{j,j'}$ are chosen such that the
matrix $\bun+\bM$ is invertible, in which case we can make the following identifications for the quantities introduced in
\eqref{eq:uijdef} and \eqref{eq:s} respectively:
\begin{equation}\label{eq:sol-uijs}
u_{i,j}=\tbs\,\bK'^j\,(\bun+\bM)^{-1}\,\bK^i\,\brr\  , \quad s_{a,b}=\tbs\,(-b\bun+\bK')^{-1}\,(\bun+\bM)^{-1}\,(a\bun+\bK)^{-1}\,\brr
\end{equation}
in which $\bK$ and $\bK'$ are the block diagonal matrices:
\bse\label{eq:KK'}\begin{eqnarray}
\bK&=&{\rm diag}\left(k_{1,1},\dots, k_{1,N_1};k_{2,1},\dots,k_{2,N_2};k_{3,1},\dots,k_{3,N_3}\right) \  , \label{eq:KK'a} \\
\bK'&=&{\rm diag}\left(-\oa_1(k_{1,1}),\dots,-\oa_1(k_{1,N_1});-\oa_2(k_{2,1}),\dots,
-\oa_2(k_{2,N_2});-\oa_3(k_{3,1}),\dots,-\oa_3(k_{3,N_3})\right)\  , \nn \\
&& \label{eq:KK'b}
\end{eqnarray}\ese
and where $\brr$ and $\tbs$ are the vectors
\bse\label{eq:rs}\begin{eqnarray}
\brr &=& {\rm diag}\left(\rho_{k_{1,1}},\dots,\rho_{k_{1,N_1}};\rho_{k_{2,1}},\dots,\rho_{k_{2,N_2}};\rho_{k_{3,1}},\dots,\rho_{k_{3,N_3}}\right)^T
\  , \label{eq:rsa} \\
\tbs &=& {\rm diag}\left(\sg_{-\oa_1(k_{1,1})},\dots,\sg_{-\oa_1(k_{1,N_1})};\sg_{-\oa_2(k_{2,1})},\dots,
\sg_{-\oa_2(k_{2,N_2})};\sg_{-\oa_3(k_{3,1})},\dots,\sg_{-\oa_3(k_{3,N_3})}\right)\  . \nn \\
&& \label{eq:rsb}
\end{eqnarray}\ese
The soliton solutions of the other relevant variables, defined in \eqref{eq:objs} are obtained in a similar way, and lead to the expressions
\bse\label{eq:sol-objs}\bea
&& v_a: = 1-\tbs\,(\bun+\bM)^{-1}\,(a\bun+\bK)^{-1}\,\brr
\quad,\quad w_b:= 1+\tbs\,(-b\bun+\bK')^{-1}\,(\bun+\bM)^{-1}\,\brr\  , \nn \\
&& s_a: = a-\tbs\,\bK'\,(\bun+\bM)^{-1}\,(a\bun+\bK)^{-1}\,\brr
\quad,\quad t_b:= -b+\tbs\,(-b\bun+\bK')^{-1}\,(\bun+\bM)^{-1}\,\bK\,\brr\  , \nn \\
&& r_a: = a^2-\tbs\,\bK'^2\,(\bun+\bM)^{-1}\,(a\bun+\bK)^{-1}\,\brr\quad,\quad
z_b:= b^2+\tbs\,(-b\bun+\bK')^{-1}\,(\bun+\bM)^{-1}\,\bK^2\,\brr\  . \nn  \eea\ese
Note that all these expressions \eqref{eq:sol-uijs} and \eqref{eq:sol-objs} can be made quite explicit after choosing the various
parameters of the solution. Finally the relevant $\tau$-function associated with these soliton solutions is given by
\begin{equation}\label{eq:tau}
\tau:=\det\left(\bun+\bM\right)
\end{equation}
and it can be shown that it obeys the following \textit{trilinear} equation:
\begin{eqnarray}\label{eq:tril}
&& (3p^2-2\alpha_2 p+\alpha_1) \wh{\tau}\tau\underaccent{\what}{\tau}+(3q^2-2\alpha_2 q+\alpha_1) \wt{\tau}\tau\underaccent{\wtilde}{\tau}
-(p-q)^2 \wh{\wt{\tau}}\tau \underaccent{\wtilde}{\underaccent{\what}{\tau}}  \nn \\
&& =[(p^2+pq+q^2)-\alpha_2(p+q)+\alpha_1]\left( \underaccent{\wtilde}{\wh{\tau}}\wt{\tau}\underaccent{\what}{\tau}
+\underaccent{\what}{\wt{\tau}}\wh{\tau}\underaccent{\wtilde}{\tau}\right)\  .
\end{eqnarray}
This trilinear equation is derived from a combination of \eqref{eq:svurels}, \eqref{eq:bA2} and \eqref{eq:papb}
together with the identifications
\be\label{eq:tau-ident}
 v_p=\frac{\underaccent{\wtilde}{\tau}}{\tau}\  , \quad w_p=\frac{\wt{\tau}}{\tau}\ , \quad v_q=\frac{\underaccent{\what}{\tau}}{\tau}
\  , \quad w_q=\frac{\wh{\tau}}{\tau}\  , \quad (q-p)S_{p,q}=\frac{\underaccent{\wtilde}{\wh{\tau}}}{\tau}\   .
\ee
The explicit expressions \eqref{eq:sol-objs} can be subsequently inserted in the formulae of the previous section identifying the
various dependent variables of the systems of \cite{H}, and thus we get the explicit multi-soliton solutions for the latter.
Furthermore, the identifications \eqref{eq:tau-ident} can be used to express these quantities in terms of the $\tau$-function.
It is to be expected that the  trilinear equation could be cast in bilinear form, but at the expense of having to introduce
additional $\tau$-functions (e.g. by introducing functions obtained by applying a shift on \eqref{eq:tau} in a lattice direction other
than the ones appearing in the equation itself), but we will not pursue that line of investigation here.

\section{Lax representations}

The relations \eqref{eq:ukrels} are the basis for the derivation of
Lax pairs for the lattice equations list in section 3. In fact, by
selecting specific components of these vectors, or combinations
thereof, we can constitute the basic vector functions in terms of
which we obtain the relevant linear problems. This the DL scheme
will provide some of the Lax pairs directly. In some cases we find
it more convenient to rely on the 3D-consistency of the systems to
produce the Lax pairs, along the lines of the papers \cite{FNij,BS},
cf. also \cite{W,TN2} for the BSQ case. \vspace{.3cm}

\noindent
{\bf \underline{Case A-2}:} \\
For the (A-2) equation \eqref{eq:bA2}, we define an eigenvector of
the form(cf. \cite{GD})
$$  \bphi=\left(\,((a+\Ld)^{-1}\bu_k)_0 ,(\bu_{k})_0,(\Ld\bu_{k})_0\right)^T. $$
Here for a infinite-component column vector $\mathbf{v}$,
$(\mathbf{v})_0$ stands for the $0$-th component of $\mathbf{v}$. In
that case one can derive the following Lax relation
~$\wt{\bphi}=\bL^{A2}\bphi$~ with: \be\label{eq:A2-L} \bL^{A2}=
\left( \begin{array}{ccccc}
p-a && \wt{v}_a && 0 \\
0 && p-\wt{u}_0 && 1 \\
\frac{G(k,-a)}{v_a} && \ast && p-\alpha_2+\frac{s_a}{v_a}
\end{array}\right)
\ee in which
$\ast=(p-\wt{u}_0)(p-\alpha_2+s_a/v_a)-p_a\wt{v}_a/v_a$, and
$\bM^{A2}$ is obtained from \eqref{eq:A2-L} by replacing $p$ by $q$
and $\wt{\phantom{a}}$ by $\wh{\phantom{a}}$. The compatibility
condition of the Lax pair, i.e. the matrix equation
$\wh{\bL^{A2}}\bM^{A2}=\wt{\bM^{A2}}\bL^{A2}$ leads to the equations
\eqref{eq:svurels_a} and \eqref{eq:svurels_b}, together with the
relation
\begin{equation}\label{eq:svvrel}
p-q+\frac{\wh{s}_a}{\wh{v}_a}-\frac{\wt{s}_a}{\wt{v}_a}
=(p-a)\frac{v_a}{\wt{v}_a}-(q-a)\frac{v_a}{\wh{v}_a}\ ,
\end{equation}
The latter relation follow readily from \eqref{eq:strels_a}.
\vspace{.3cm}

\noindent
{\bf \underline{Case B-2}:} \\

For the extended lattice BSQ equation \eqref{eq:u00_ab} and
\eqref{eq:BSQ-eq3}, then one can define vector
$$\bpsi=((\bu_{k})_0,(\Ld\bu_{k})_0,(\oa_1(-\Ld)\oa_2(-\Ld)\bu_{k})_0)^T,$$
then from \eqref{eq:ukrels},  we obtain \bse\label{eq:BSQ-LMk}\bea
&& \wt{\bpsi}=\bL^{BSQ}\bphi \ , \\
&& \wh{\bpsi}=\bM^{BSQ}\bphi \ , \eea\ese where \be\label{eq:BSQ-Lk}
\bL^{BSQ}=\left(
\begin{array}{ccc}
p-\wt{u}_0 & 1 & 0\\
-\wt{u}_{1,0}-\alpha_1 & p+\alpha_2 & 1 \\
\ast & -u_{0,1}+2\alpha_2u_0-2\alpha_2^2& p-2\alpha_2+u_0
\end{array}\right),
\ee in which
$\ast=G(k,-p)-(p-\wt{u}_0)[(p-\alpha_2)(p+u_0)+u_{0,1}]-(p+u_0-2\alpha_2)(\wt{u}_{1,0}+\alpha_1)$,
and where $\bM^{BSQ}$ is obtained from \eqref{eq:BSQ-Lk} by
replacing $p$ by $q$ and $\wt{\phantom{a}}$ by $\wh{\phantom{a}}$.
The compatibility
$\wh{\bL^{BSQ}}\bM^{BSQ}=\wt{\bM^{BSQ}}\bL^{BSQ}$ leads to the
equations \bse\be
\wh{u}_{1,0}-\wt{u}_{1,0}=(p-q+\wh{u}_{0}-\wt{u}_{0})\wh{\wt{u}}_{0}-p\wh{u}_{0}+q\wt{u}_{0}\
\label{eq:u012-h-t_b} \ee
and
\be\label{eq:u01}
\wh{u}_{0,1}-\wt{u}_{0,1}=(p-q+\wh{u}_0-\wt{u}_0)u_0-p\wt{u}_0+q\wh{u}_0,
\ee\ese together with eq. \eqref{eq:BSQ-eq3}. These three equations
constitute a system which is equivalent to the extended lattice BSQ
equation \eqref{eq:u00_ab}  and \eqref{eq:BSQ-eq3}. \vspace{.3cm}

In the following cases, we give Lax pairs which are derived on the
basis of the 3D consistency property for the systems of three coupled
equation which we derived in section 4. In the generic case
for a 3-component system containing variables $(x,y,z)$, we would need to set $(\wb x=g/f, ~\wb
y=h/f,~ \wb z=k/f)$ and constitute the vector is $\phi=(g,h,k,f)^T$ leading
to a Lax pair in terms of $4\times 4$ matrices (see, e.g.,  \eqref{eq:lax-4}
and \eqref{eq:lax-4-c4} below). However in some cases the $4\times 4$ situation
simplifies to a Lax pair in terms of $3\times 3$ matrices for the 3-component system
(e.g., in the cases of the A-2 and B-2 systems).

\vspace{.3cm}

\noindent
{\bf \underline{Case A-2} revisited:} \\
For the extended three-component lattice equation \eqref{eq:A2-ex},
we extend the lattice equation into a third dimension by introducing
a new variable $l$ associated with a new shift $\wb{\phantom{a}}$
and a new complex parameter $r$ \bse\label{eq:A2-ex-3DC} \bea
&& \wt{y} = z\wt{x}-x \ ,~~\wh{y} = z\wh{x}-x \ ,~~\wb{y} = z\wb{x}-x\  , \label{eq:A2-ex-3DC_a}\\
&& y =x\wh{\wt{z}}-b_0 x+\frac{-G(-p,-a)\ \wt{x}+G(-q,-a)\,\wh{x}}{\wh{z}-\wt{z}}\ ,\label{eq:A2-ex-3DC_b1}\\
&& y =x\wb{\wt{z}}-b_0 x+\frac{-G(-p,-a)\ \wt{x}+G(-r,-a)\,\wb{x}}{\wb{z}-\wt{z}}\ ,\label{eq:A2-ex-3DC_b2}\\
&& y =x\wh{\wb{z}}-b_0 x+\frac{-G(-r,-a)\
\wb{x}+G(-q,-a)\,\wh{x}}{\wh{z}-\wb{z}}\ ,\label{eq:A2-ex-3DC_b3}
\eea\ese
From \eqref{eq:A2-ex-3DC_a}, it is easy to derive \bse
\label{eq:A2-com}\bea \wh{\wt{x}} & = &
\frac{\wt{x}-\wh{x}}{\wt{z}-\wh{z}} \ ,~~\wt{\wb{x}} =
\frac{\wt{x}-\wb{x}}{\wt{z}-\wb{z}} \ ,~~\wh{\wb{x}} =
\frac{\wb{x}-\wh{x}}{\wb{z}-\wh{z}} \ , \label{eq:A2-com-a} \\
\wh{\wt{y}}& = & \frac{\wt{x}\wh{z}-\wh{x}\wt{z}}{\wt{z}-\wh{z}} \
,~~ \wt{\wb{y}} = \frac{\wt{x}\wb{z}-\wb{x}\wt{z}}{\wt{z}-\wb{z}}  \
,~~ \wh{\wb{y}} = \frac{\wb{x}\wh{z}-\wh{x}\wb{z}}{\wb{z}-\wh{z}} \
. \label{eq:A2-com-b}
 \eea\ese
By the 3D-consistency, setting
$\bar{x}=\frac{\varphi_1}{\varphi_0},~\bar{z}=\frac{\varphi_2}{\varphi_0},~\bar{y}=\frac{\varphi_3}{\varphi_0}$,
then the Lax pair is \be\label{eq:A2-LP}
\wt{\bphi}_{11}=\bL_{11}\bphi_{11} \ ,
~~\wh{\bphi}_{11}=\bM_{11}\bphi_{11} \ , \ee in which \be
\bphi_{11}=\left(\begin{array}{c}\varphi_0\\\varphi_1\\\varphi_2\\\varphi_3
\end{array}\right),~~
\bL_{11}=\left(
\begin{array}{cccc}
 \wt{z} & 0 & -1& 0 \\
\wt{x} &  -1  & 0 & 0\\
(\frac{y}{x}+b_0)\wt{z}-\frac{G(-p,-a)\wt{x}}{x} & \frac{G(-r,-a)}{x}& -(\frac{y}{x}+b_0) & 0  \\
0 &  -\wt{z} & \wt{x}& 0
\end{array}\right),
\ee where the matrix $\bM_{11}$ is the hat-$q$ version of
$\bL_{11}$. The compatibility gives the relations:
\begin{equation}\label{eq:C3compat}
\wh{\wt x}=\frac{\wt x-\wh x}{\wt z-\wh z}\  , \qquad x=\frac{\wh
x\wt y-\wt x\wh y}{\wt x-\wh x}\  ,
\end{equation}
together with eq. \eqref{eq:4.3b}. The two equations
\eqref{eq:C3compat} are consequences of the 3-component system which
constitutes the case (A-2). We note that there is a zero column
appearing in $\bL_{11}$ as well as in $\bM_{11}$, which indicates
$\wb y$ is not necessary to be used to generate a Lax pair. Let us
remove the last columns and rows in $\bL_{11}$ and $\bM_{11}$, and
examine the remains: \be\label{eq:A2-LP}
\wt{\bphi}_{12}=\bL_{12}\bphi_{12},
~~\wh{\bphi}_{12}=\bM_{12}\bphi_{12}, \ee in which \be
\bphi_{12}=\left(\begin{array}{c}\varphi_0\\\varphi_1\\\varphi_2
\end{array}\right),~~
\bL_{12}=\left(
\begin{array}{ccc}
 \wt{z} & 0 & -1\\
\wt{x} &  -1 & 0 \\
(\frac{y}{x}+b_0)\wt{z}-\frac{G(-p,-a)\wt{x}}{x} & \frac{G(-r,-a)}{x} & -(\frac{y}{x}+b_0) \\
\end{array}\right),
\ee where the matrix $\bM_{12}$ is the hat-$q$ version of
$\bL_{12}$. The compatibility of \eqref{eq:A2-LP} gives
\eqref{eq:C3compat} and \eqref{eq:4.3b} as well.

\vspace{.3cm}

\noindent
{\bf \underline{Case B-2} revisited:} \\
Similarly, for the extended three-component lattice BSQ equation
\eqref{eq:BSQ-ex}, by considering the third dimension of this
lattice equation, then we have \bse\label{eq:BSQ-ex-3DC}\bea
&& \wt{z} = x\wt{x}-y,~~\wh{z} = x\wh{x}-y,~~\wb{z} = x\wb{x}-y\ ,\label{eq:BSQ-ex-3DC-a} \\
&& z = x\wh{\wt{x}}-\wh{\wt{y}}-\alpha_2(\wh{\wt{x}}-x)-\alpha_1-\frac{G(-p,-q)}{\wh{x}-\wt{x}}\ , \label{eq:BSQ-ex-3DC-b1}\\
&& z = x\wb{\wt{x}}-\wb{\wt{y}}-\alpha_2(\wb{\wt{x}}-x)-\alpha_1-\frac{G(-p,-r)}{\wb{x}-\wt{x}}\ , \label{eq:BSQ-ex-3DC-b2}\\
&& z =
x\wh{\wb{x}}-\wh{\wb{y}}-\alpha_2(\wh{\wb{x}}-x)-\alpha_1-\frac{G(-r,-q)}{\wh{x}-\wb{x}}\
. \label{eq:BSQ-ex-3DC-b3} \eea\ese From \eqref{eq:BSQ-ex-3DC-a}, it
is easy to derive \bse\label{eq:BSQ-com}\bea \wh{\wt{x}} & = &
\frac{\wt{y}-\wh{y}}{\wt{x}-\wh{x}} \ ,~~\wb{\wt{x}} =
\frac{\wt{y}-\wb{y}}{\wt{x}-\wb{x}} \ ,
~~\wh{\wb{x}} = \frac{\wb{y}-\wh{y}}{\wb{z}-\wh{x}} \ , \label{eq:BSQ-com-a} \\
\wh{\wt{z}} & = & \frac{\wh{x}\wt{y}-\wt{x}\wh{y}}{\wt{x}-\wh{x}}  \
,~~ \wh{\wb{z}} = \frac{\wh{x}\wb{y}-\wb{x}\wh{y}}{\wb{x}-\wh{x}} \
,~~ \wt{\wb{z}} = \frac{\wt{x}\wb{y}-\wb{x}\wt{y}}{\wb{x}-\wt{x}} \
. \label{eq:BSQ-com-B} \eea\ese By the 3D-consistency, setting
$\bar{x}=\frac{\phi_1}{\phi_0},~\bar{y}=\frac{\phi_2}{\phi_0},~\bar{z}=\frac{\phi_3}{\phi_0}$,
then the Lax pair is \be\label{eq:BSQ-LP}
\wt{\bphi}_{21}=\bL_{21}\bphi_{21} \ ,
~~\wh{\bphi}_{21}=\bM_{21}\bphi_{21} \ , \ee in which \be
\bphi_{21}=\left(\begin{array}{c}\phi_0\\ \phi_1 \\ \phi_2 \\
\phi_3\end{array}\right),~~ \bL_{21}=\left(
\begin{array}{cccc}
\wt{x} & -1 & 0 & 0\\
\wt{y} &  0 & -1 & 0\\
\ast & z-\alpha_2x+\alpha_1 & \alpha_2 -x & 0 \\
0 &  \wt{y} & -\wt{x} & 0\\
\end{array}\right) \ \label{eq:BSQ-L21},
\ee with $\ast=-(z-\alpha_2
x+\alpha_1)\wt{x}+\wt{y}(x-\alpha_2)+G(-p,-r)$, and where the matrix
$\bM_{21}$ is the hat-$q$ version of $\bL_{21}$. The compatibility
yields the relations
\begin{equation}\label{eq:xompB3}
\wh{\wt{x}} = \frac{\wt{y}-\wh{y}}{\wt{x}-\wh{x}}\  ,\qquad x=
\frac{\wt{z}-\wh{z}}{\wt{x}-\wh{x}}
\end{equation}
together with the relation \eqref{eq:4.9b}. Similar to the case
(A-2) the Lax matrices $\bL_{21}$ and $\bM_{21}$ can degenerate to
$3 \times 3$ case by removing their last rows and columns. The
results are \be\label{eq:BSQ-LP} \wt{\bphi}_{22}=\bL_{22}\bphi_{22}
\ , ~~\wh{\bphi}_{22}=\bM_{22}\bphi_{22} \ , \ee in which \be
\bphi_{22}=\left(\begin{array}{c}\phi_0\\ \phi_1 \\
\phi_2\end{array}\right) \ ,~~ \bL_{22}=\left(
\begin{array}{ccc}
\wt{x} & -1 & 0\\
\wt{y} &  0 & -1 \\
\ast & z-\alpha_2x+\alpha_1 & \alpha_2 -x \\
\end{array}\right) \ ,
\ee with $\ast$ is defined as in \eqref{eq:BSQ-L21} and the matrix
$\bM_{22}$ is the hat-$q$ version of $\bL_{22}$. The compatibility
yields the relations \eqref{eq:xompB3} and \eqref{eq:4.9b}

\vspace{.3cm}

\noindent
{\bf \underline{Case C-3}:} \\
For this case, which includes the extended three-component lattice
MBSQ/SBSQ equation \eqref{eq:xyz-MSBSQ}, we are led by considering a
third dimension of this lattice equation to the following relations:
\bse\label{eq:xyz-MSBSQ-3DC}\bea && x-\wt{x} = \wt{y}z\  , ~~
x-\wh{x} = \wh{y}z\  ,  ~~
x-\wb{x} = \wb{y}z\  ,\label{eq:xyz-MSBSQ-3DC-a} \\
&& y \wh{\wt{z}} = z\frac{-G(-p,-b)\wh{z}\wt{y}+G(-q,-b)\wt{z}\wh{y}}{\wt{z}-\wh{z}}+G(-a,-b)\wh{\wt{x}}\ ,  \label{eq:xyz-MSBSQ-3DC-b1}\\
&& y \wb{\wt{z}} = z\frac{-G(-p,-b)\wb{z}\wt{y}+G(-r,-b)\wt{z}\wb{y}}{\wt{z}-\wb{z}}+G(-a,-b)\wb{\wt{x}}\ , \label{eq:xyz-MSBSQ-3DC-b2}\\
&& y \wh{\wb{z}} =
z\frac{-G(-r,-b)\wh{z}\wb{y}+G(-q,-b)\wb{z}\wh{y}}{\wb{z}-\wh{z}}+G(-a,-b)\wh{\wb{x}}.
\label{eq:xyz-MSBSQ-3DC-b3} \eea\ese From
\eqref{eq:xyz-MSBSQ-3DC-a}, it is easy to derive
\bse\label{eq:xyz-MSBSQ-3DC-a-C}\bea && \wh{\wt{x}} =
\frac{\wh{x}\wt{z}-\wt{x}\wh{z}}{\wt{z}-\wh{z}}\ ,~~ \wb{\wt{x}} =
\frac{\wb{x}\wt{z}-\wt{x}\wb{z}}{\wt{z}-\wb{z}}\ , ~~
\wh{\wb{x}} = \frac{\wh{x}\wb{z}-\wb{x}\wh{z}}{\wb{z}-\wh{z}} ,\\
&& \wh{\wt{y}} =
\frac{\wt{x}-\wh{x}}{\wt{z}-\wh{z}}=-z\frac{\wt{y}-\wh{y}}{\wt{z}-\wh{z}}\
,~~ \wb{\wt{y}} =
\frac{\wt{x}-\wb{x}}{\wt{z}-\wb{z}}=-z\frac{\wt{y}-\wb{y}}{\wt{z}-\wb{z}}\
,~~ \wh{\wb{y}} =
\frac{\wb{x}-\wh{x}}{\wb{z}-\wh{z}}=-z\frac{\wb{y}-\wh{y}}{\wb{z}-\wh{z}}\
. \eea\ese By the 3D-consistency, setting
$\bar{x}=\frac{\psi_1}{\psi_0},~\bar{y}=\frac{\psi_2}{\psi_0},~\bar{z}=\frac{\psi_3}{\psi_0}$,
then we have the Lax pair \be\label{eq:MSBSQ-LP}
\wt{\bpsi}_1=\bL_3\bpsi_1, ~~\wh{\bpsi}_1=\bM_3\bpsi_1, \ee in which
$\bpsi_1=(\psi_0,~\psi_1,~\psi_2,~\psi_3)^T$ and \be
\bL_3=\frac{1}{z}\left(
\begin{array}{cccc}
 -\wt{z} & 0 & 0 & 1 \\
0 &  -\wt{z} & 0 & \wt{x} \\
z\wt{y} &  0 & -z & 0 \\
0 &  -G(-a,-b)\frac{\wt{z}}{y} & -G(-r,-b)\frac{z\wt{z}}{y} &
G(-p,-b)\frac{z\wt{y}}{y}+G(-a,-b)\frac{\wt{x}}{y}
\end{array}\right),
\label{eq:lax-4} \ee where the matrix $\bM_3$ is the hat-$q$ version
of $\bL_3$. The compatibility yields now:
\begin{equation}
 \wh{\wt{x}} = \frac{\wh{x}\wt{z}-\wt{x}\wh{z}}{\wt{z}-\wh{z}}\ , \qquad
 \wh{\wt{y}} = -z\frac{\wt{y}-\wh{y}}{\wt{z}-\wh{z}}
 \end{equation}
together with the relation \eqref{eq:xyz-MSBSQ-b}. They constitute
the case (C-3) of the extended BSQ family. \vspace{.3cm}

\noindent
{\bf \underline{Alternate Case C-3}:} \\
In a similar way as in the previous case, for the extended
three-component lattice MBSQ/SBSQ equation \eqref{eq:xyz-MSBSQ-1},
by considering a  third dimension where we impose this lattice
system, we have \bse\label{eq:xyz-MSBSQ-1-3DC}\bea && x-\wt{x} =
\wt{y}z\  , ~~ x-\wh{x} = \wh{y}z\  ,  ~~
x-\wb{x} = \wb{y}z\  ,\label{eq:xyz-MSBSQ-1-3DC-a} \\
&& y \wh{\wt{z}} = z\frac{-G(-p,-a)\wh{z}\wt{y}+G(-q,-a)\wt{z}\wh{y}}{\wt{z}-\wh{z}}+G(-a,-b)x\ ,  \label{eq:xyz-MSBSQ-1-3DC-b1}\\
&& y \wb{\wt{z}} = z\frac{-G(-p,-a)\wb{z}\wt{y}+G(-r,-a)\wt{z}\wb{y}}{\wt{z}-\wb{z}}+G(-a,-b)x\ , \label{eq:xyz-MSBSQ-1-3DC-b2}\\
&& y \wh{\wb{z}} =
z\frac{-G(-r,-a)\wh{z}\wb{y}+G(-q,-a)\wb{z}\wh{y}}{\wb{z}-\wh{z}}+G(-a,-b)x\
. \label{eq:xyz-MSBSQ-1-3DC-b3} \eea\ese Obviously, the identities
\eqref{eq:xyz-MSBSQ-3DC-a-C} are still hold. By using
3D-consistency, and setting
$\bar{x}=\frac{\mu_1}{\mu_0},~\bar{y}=\frac{\mu_2}{\mu_0},~\bar{z}=\frac{\mu_3}{\mu_0}$,
then from \eqref{eq:xyz-MSBSQ-3DC-a-C},
\eqref{eq:xyz-MSBSQ-1-3DC-b2} and \eqref{eq:xyz-MSBSQ-1-3DC-b3} we
have the Lax pair \be\label{eq:MSBSQ-1-LP}
\wt{\bpsi}_{2}=\bL_{4}\bpsi_{2}, ~~\wh{\bpsi}_{2}=\bM_{4}\bpsi_{2},
\ee in which $\bpsi_{2}=(\mu_0,~\mu_1,~\mu_2,~\mu_3)^T$ and \be
\bL_{4}= \frac{1}{z}\left(
\begin{array}{cccc}
 -\wt{z} & 0 & 0 & 1 \\
0 &  -\wt{z} & 0 & \wt{x} \\
z\wt{y} &  0 & -z & 0 \\
-G(-a,-b)\frac{\wt{z}x}{y}  & 0 &  -G(-r,-a)\frac{z\wt{z}}{y} & G(-p,-a)\frac{z\wt{y}}{y}+G(-a,-b)\frac{x}{y} \\
\end{array}\right),
\ee where the matrix $\bM_{4}$ is the hat-$q$ version of $\bL_{4}$.
The compatibility now produces the relations
\begin{equation} \label{eq:altC3}
\wh{\wt{x}} = \frac{\wh{x}\wt{z}-\wt{x}\wh{z}}{\wt{z}-\wh{z}}\ ,~~x
= \frac{\wh{x}\wt{y}-\wt{x}\wh{y}}{\wt{y}-\wh{y}} \ , ~~
 \wh{\wt{y}} = -z\frac{\wt{y}-\wh{y}}{\wt{z}-\wh{z}}
\end{equation}
together with \eqref{eq:xyz-MSBSQ-1b}. Once again this system is a
consequence of the original lattice system.

\vspace{.3cm}

\noindent
{\bf \underline{Case C-4}:} \\
For this case, which includes the extended three-component lattice
MBSQ/SBSQ equation \eqref{eq:C4-a} and \eqref{eq:C4-b}, we have the
following 3D-consistent relations:
\bse\label{eq:xyz-MSBSQ-1-3DC}\bea && x_1-\wt{x}_1 = \wt{y}_1z_1\  ,
~~ x_1-\wh{x}_1 = \wh{y}_1z_1\  ,  ~~
x_1-\wb{x}_1 = \wb{y}_1z_1\  ,\label{eq:xyz-MSBSQ-1-3DC-a} \\
&& y_1 \wh{\wt
z}_1=z_1\,\frac{P_{a,b}\,\wh{z}_1\wt{y}_1-Q_{a,b}\,\wt{z}_1\wh{y}_1}{\wt{z}_1-\wh{z}_1}+
x_1 \wh{\wt x}_1-G^2_{a,b}\ ,
\label{eq:xyz-MSBSQ-1-3DC-b1}\\
&& y_1 \ol{\wt
z}_1=z_1\,\frac{P_{a,b}\,\ol{z}_1\wt{y}_1-R_{a,b}\,\wt{z}_1\ol{y}_1}{\wt{z}_1-\ol{z}_1}+
x_1 \ol{\wt x}_1-G^2_{a,b}\ ,
\label{eq:xyz-MSBSQ-1-3DC-b2}\\
&& y_1 \wh{\ol
z}_1=z_1\,\frac{R_{a,b}\,\wh{z}_1\ol{y}_1-Q_{a,b}\,\ol{z}_1\wh{y}_1}{\ol{z}_1-\wh{z}_1}+
x_1 \wh{\ol x}_1-G^2_{a,b}, \label{eq:xyz-MSBSQ-1-3DC-b3} \eea\ese
where $R_{a,b}=-\frac{1}{2}(G(-r,-b)+G(-r,-a))$. From
\eqref{eq:xyz-MSBSQ-1-3DC-a} we have same consistent forms as
\eqref{eq:xyz-MSBSQ-3DC-a-C} with $(x,y,z)$ replaced by
$(x_1,y_1,z_1)$. Setting
$\wb{x}_1=\frac{\nu_1}{\nu_0},~\wb{y}_1=\frac{\nu_2}{\nu_0},~\wb{z}_1=\frac{\nu_3}{\nu_0}$,
we have the Lax pair \be\label{eq:MSBSQ-LP}
\wt{\bpsi}_3=\bL_5\bpsi_3, ~~\wh{\bpsi}_3=\bM_5\bpsi_3, \ee in which
$\bpsi_3=(\nu_0,~\nu_1,~\nu_2,~\nu_3)^T$ and \be
\bL_5=\frac{1}{z_1}\left(
\begin{array}{cccc}
 -\wt{z}_1 & 0 & 0 & 1 \\
0 &  -\wt{z}_1 & 0 & \wt{x}_1 \\
z_1\wt{y}_1 &  0 & -z_1 & 0 \\
G^2_{a,b}\frac{\wt{z}_1}{y_1} &  -\frac{x_1\wt{z}_1}{y_1} &
R_{a,b}\frac{z_1\wt{z}_1}{y_1} &
-P_{a,b}\frac{\wt{y}_1z_1}{y_1}+\frac{x_1\wt{x}_1}{y_1}-\frac{G^2_{a,b}}{y_1}
\end{array}\right),
\label{eq:lax-4-c4} \ee where the matrix $\bM_5$ is the hat-$q$
version of $\bL_5$. Now the compatibility yields
\begin{equation}
  x_1 = \frac{\wh{x}_1\wt{y}_1-\wt{x}_1\wh{y}_1}{\wt{y}_1-\wh{y}_1}\ ,\qquad
  \wh{\wt{x}}_1 = \frac{\wh{x}_1\wt{z}_1-\wt{x}_1\wh{z}_1}{\wt{z}_1-\wh{z}_1}\ , \qquad
 \wh{\wt{y}}_1 = -z_1\frac{\wt{y}_1-\wh{y}_1}{\wt{z}_1-\wh{z}_1}
 \end{equation}
together with the relation \eqref{eq:C4-b}. They constitute the case
(C-4) of the extended BSQ family.

\section{Conclusions}

Since the realization that the notion of multidimensional
consistency is a key criterion for integrability of partial
difference equations, the search for new integrable equations on the
2D lattice using this notion as the defining property has led to
many novel results. The ABS classification, \cite{ABS}, of
single-variable quadrilateral lattice equations has demonstrated
that such systems are highly restricted (up to equivalence), but the
non-scalar case remains still to be largely explored. The search in
\cite{H} of 2-and 3-component systems of a form inspired by earlier
results from the BSQ family of lattice equations, \cite{GD,DIGP,TN2}
seems to confirm the assertion that the realm of such discrete
equations has only few members, but the examples are rich and
important. The parameter-extensions of the BSQ systems established
in \cite{H} are noteworthy because the extra parameter freedom
allows their soliton solutions to exhibit some new features, cf.
\cite{HZ,HZh}. On the other hand, the ``search methodology''
exploited in \cite{H} does not reveal all there is to know about
these equations, and this is supplemented by the structural approach
wich we took in the present paper. Thus, by extending the DL
approach of \cite{GD} to include more general types of dispersion
relations, we have been able in the present work to identify all the
variables and their interconnections of the extended BSQ systems of
\cite{H}. In fact, this amounts to establishing Miura type (i.e.,
non-auto B\"acklund) transformations between these various systems
(A-2), (B-3), (C-3) and (C-4), which involve also point
transformations. Furthermore, the DL structure can be used
effectively to yield explicit multi-soliton type solutions for all
these systems by specifying in an appropriate way the contours and
integration measures which define some of the main objects (e.g., in
\eqref{eq:bC}). For more general measures and contours the DL also
can yield wider classes of solutions, such as inverse scattering
type solutions or scaling reductions. To explore further, and make
explicit, the latter line of work on the basis of the present
results, will be the subject of future investigations.

\subsection*{Acknowledgements}

This project is supported by the NSF of China (No. 11071157),
Shanghai Leading Academic Discipline Project (No. J50101) and
Postgraduate Innovation Foundation of Shanghai University (No.
SHUCX111027). One of the authors (FWN) is grateful to the University
of Shanghai for its hospitality during a visit when this work was
initiated. He is currently supported by a Royal Society/Leverhulme Trust 
Senior Research Fellowship.

\section*{Appendix}
\setcounter{equation}{0}

In this Appendix we derive some integral formulae
from the general abstract structure displayed in section 2.
Starting from  \eqref{eq:bC}, i.e.,
\be\label{eq:bC-tem}
\bC= \sum_{j=1}^N
\int_{\Gamma_j}\,d\ld_j(k)\,\rho_k
\bc_k\,\bc^t_{-\oa_j(k)}\sg_{-\oa_j(k)}\ , \ee
and the defining relation for the vectors $\bu_k$, i.e.,
\be
\label{eq:uk-def-tem}
\bu_k+\rho_k\bU\,\bOm\,\bc_k=\rho_k\bc_k \ ,
\ee
we can show that
\be\label{eq:bU-tem1}
\bU= \sum_{j=1}^N
\int_{\Gamma_j}\,d\ld_j(k)\,\bu_k\,\bc^t_{-\oa_j(k)}\sg_{-\oa_j(k)}
\ee
is equivalent to the defining relation
\be\label{eq:bu-tem2}
\bU=\bC-\bU\,\bOm\,\bC.
\ee
In fact,
\[\eqref{eq:bU-tem1}-\eqref{eq:bC-tem}=\bU-\bC=
 \sum_{j=1}^N
\int_{\Gamma_j}\,d\ld_j(k)\,(\bu_k-\rho_k\bc_k)\,\bc^t_{-\oa_j(k)}\sg_{-\oa_j(k)}\ .
\]
Using \eqref{eq:uk-def-tem} it is then
\begin{align*}
\bU-\bC& =
-\bU\,\bOm\sum_{j=1}^N \int_{\Gamma_j}\,d\ld_j(k)\,\rho_k\,\bc_k\,\bc^t_{-\oa_j(k)}\sg_{-\oa_j(k)}\\
&=-\bU\,\bOm\,\bC.
\end{align*}
On the other hand,
starting from \eqref{eq:bC-tem} and using \eqref{eq:uk-def-tem}, one has
\begin{align}
\bC
&=\sum_{j=1}^N \int_{\Gamma_j}\,d\ld_j(k)\,\bu_k\,\bc^t_{-\oa_j(k)}\sg_{-\oa_j(k)}
+\bU\,\bOm\sum_{j=1}^N \int_{\Gamma_j}\,d\ld_j(k)\,\rho_k\,\bc_k\,\bc^t_{-\oa_j(k)}\sg_{-\oa_j(k)}\nonumber\\
&=\sum_{j=1}^N \int_{\Gamma_j}\,d\ld_j(k)\,\bu_k\,\bc^t_{-\oa_j(k)}\sg_{-\oa_j(k)}+\bU\,\bOm\,\bC,
\label{eq:A-bC}
\end{align}
which yields \eqref{eq:bU-tem1} if \eqref{eq:bu-tem2} holds.

Then, quite similar to the derivation in \eqref{eq:A-bC},
if starting from \eqref{eq:Csoliton-1}, i.e.,
\begin{equation}\label{eq:A-Csoliton-1}
\bC= \sum_{j=1}^3 \sum_{j'=1}^{N_j} \Lambda_{j,j'}
\rho_{k_{j,j'}}\,
\bc_{k_{j,j'}}\,\tbc_{-\oa_j(k_{j,j'})}\sg_{-\oa_j(k_{j,j'})}\ ,
\end{equation}
and using \eqref{eq:uk-def-tem} and \eqref{eq:bu-tem2}, we immediately have
\begin{equation}\label{eq:Usoliton-1}
\bU= \sum_{j=1}^3 \sum_{j'=1}^{N_j} \Lambda_{j,j'}
\bu_{k_{j,j'}}\,\tbc_{-\oa_j(k_{j,j'})} \sg_{-\oa_j(k_{j,j'})} \ ,
\end{equation}
which is \eqref{eq:Usoliton-1}.
Finally, for $\bu_k$ we have
\begin{align*}
\bu_k- \rho_k \bc_k &= -\rho_k \bU\,\bOm\,\bc_k\\
&= -\rho_k \sum_{j=1}^3 \sum_{j'=1}^{N_j} \Lambda_{j,j'}\bu_{k_{j,j'}}\tbc_{-\oa_j(k_{j,j'})}\,\frac{k-\oa_j(k_{j,j'})} {k-\oa_j(k_{j,j'})}\,
 \bOm\,\bc_{k}\, \sg_{-\oa_j(k_{j,j'})}\\
&= -\rho_k \sum_{j=1}^3 \sum_{j'=1}^{N_j} \Lambda_{j,j'}\bu_{k_{j,j'}}\,\frac{\sg_{-\oa_j(k_{j,j'})}}{k-\oa_j(k_{j,j'})}\ ,
\end{align*}
which gives \eqref{eq:uksoliton-1},
where we have successfully made use of
\[ \Ld\,\bc_k=k\,\bc_k ,\quad
\bc^t_{k'}\tLd=k'\,\bc^t_{k'},~~
\tLd \bOm + \bOm \Ld=\bO,~~\bO=\bme\,\tbme\]
and $\bc^t_{k'}\,\bO\,\bc_k=1$.

\vskip 15pt
\end{document}